\documentclass[11pt]{article}

\usepackage{graphicx}
\usepackage{amsmath}
\usepackage{amssymb,amsmath,wasysym}
\usepackage[a4paper,left=20mm,right=20mm,top=15mm,bottom=15mm]{geometry}

\usepackage[utf8]{inputenc}
\usepackage[english]{babel}
\begin{document}

\title{\bf On the speed of sound and heat capacity of liquid neon\\ in the subcritical region}

\author{A.L. Goncharov$^{1}$, V.V. Melent'ev$^{2}$, E.B. Postnikov$^{1*}$}

\date{\it $^{1}$Department of Theoretical Physics, Kursk State University, Radishcheva st., 33, 305000 Kursk, Russia\\
$^{2}$ Laboratory of Molecular Acoustics, Kursk State University, Radishcheva st., 33, 305000 Kursk, Russia\\
$^{*}$ Corresponding author. E-mail: postnicov@gmail.com}

\maketitle

\begin{abstract}
The data (the speed of sound, the isobaric and isochoric heat capacities as well as the heat capacity ratio) for liquid neon presented in NIST Chemistry WebBook are analyzed. It has been shown, basing on the representation of the inverse reduced volume fluctuations, that they consist of sufficient discrepancies in the subcritical region. The correction of data in this region of the coexistence curve is evaluated using the fluctuation approach and the theory of thermodynamic similarity.
\end{abstract}

\section{Introduction}

Liquid neon finds applications in a wide range of cryogenic technologies since it has an advantage of a larger refrigerating capacity over other noble gases or liquid hydrogen \cite{Jha2005book,Haussinger2002}. Additionally, one of the recent most promising applications is its usage in low energy particle detectors, in particular, for the observation of neutrinos \cite{Lippincott2012} and for the search of the hypothetical particles of dark matter (WIMP). This is based on the fact of neon’s better sensitivity in comparison with liquid argon and xenon, the review of recent works can be found in \cite{Lippincott2010diss,Chepel2013}.

At the same time, the thermodynamic data for this liquid are not so extensively studied as those for other simple liquids, e.g. argon, which is in fact a standard for this kind of matter. This especially regards to the state parameters at high (for neon) temperatures, i.e. in the subcritical region (the critical temperature of neon is $T_c=44.4$~K). A number of the recently accepted practically applicable approximations of experimental data could be found in the works \cite{Rabinovich1988book,Katti1986}. The fits proposed in the latter are also implemented in the modern standard reference online database: \texttt{NIST Chemistry WebBook} \cite{NIST}.

The recent study of the volume fluctuations in liquid noble gases and the corresponding coexisting vapours under saturation conditions \cite{Goncharov2013} has detected a certain irregularity in the plot of fluctuations in liquid neon in the subcritical region. Thus, the main goal of this work is to explore in details the values of thermodynamic quantities (the speed of sound, the isobaric and isochoric heat capacities) provided by \texttt{NIST Chemistry WebBook} in the direct vicinity of the critical point and to discuss their more accurate approximations within this region.

\section{Method and results}

The inverse ratio of the volume fluctuations in a condensed medium to its value for the hypothetical case of ideal gas at the same thermodynamic conditions \cite{Goncharov2013} reads as follows
\begin{equation}
\nu=\left[\left.\frac{\left\langle (\Delta V)^2\right\rangle}{V}\right/\frac{\left\langle (\Delta V)^2_{ig}\right\rangle}{V_{ig}}\right]^{-1}=\frac{\mu c^2}{\gamma RT},
\label{nu}
\end{equation}
where $\mu$, $c$, $\gamma$, $R$, $T$ are the molar mass, the speed of sound, heat capacity ratio, gas constant, and temperature. The ratio (\ref{nu}) is a very sensitive parameter, which allows for exploring the microscopic structural characteristics of fluids basing on the macroscopically measurable quantities referring on the deviation of the latter from unity.
It should be pointed out also a definite correlation of such an approach with the method of classic ideal curves on the thermodynamic surface \cite{Nedostup2015} and the correlations between the derivatives of thermodynamic functions in the subcritical region \cite{Trotsenko2013}. 

\begin{figure}
\includegraphics[width=\textwidth]{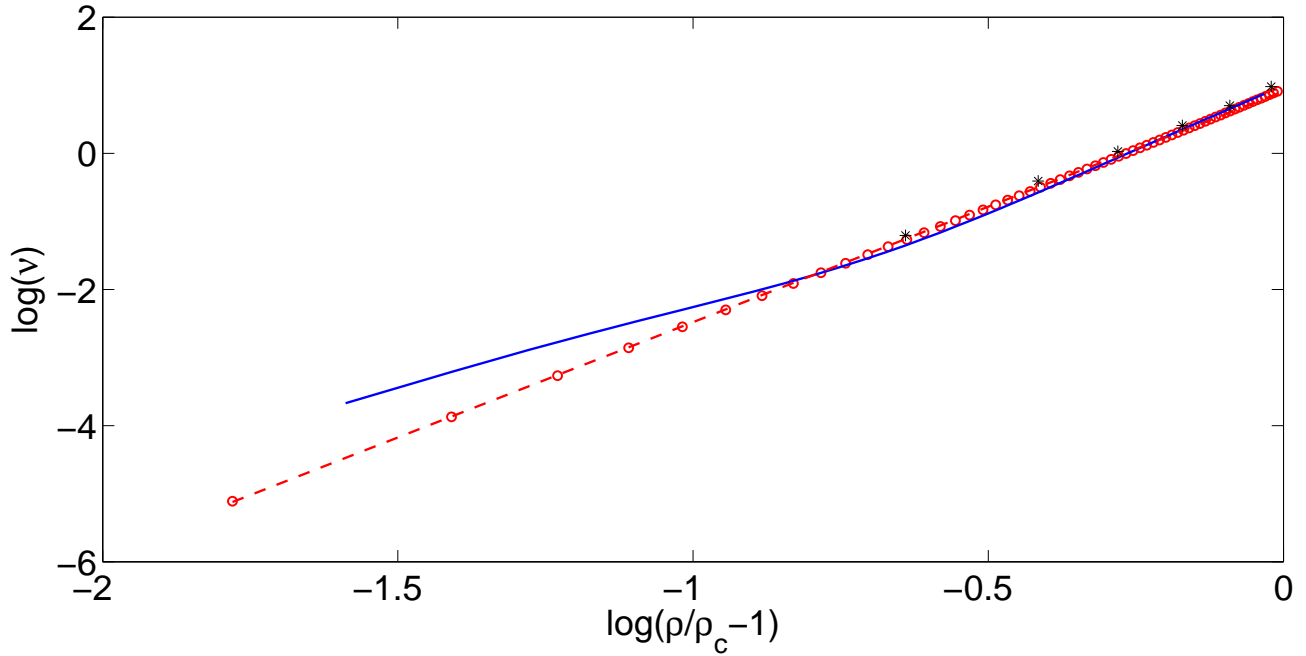}
\caption{The reduced volume fluctuations in liquid neon (solid line is drawn basing on NIST data; asterisks corresponds to the data taken from \cite{Gladun1966}) and argon (NIST-based calculations are marked with circles; dashed line demonstrates their approximation by the function $\log \nu_{Ar}=3.39\log(\rho_r-1)+0.91$).
}
\label{nufig}
\end{figure}

For the simple liquids, the dependence of the parameter $\nu$ on the density along the coexistence curve varies from the exponential one in the region close to a melting point to the fractional power law under subcritical conditions. The example of the latter is presented in Fig.~\ref{nufig} for liquid argon as a function of the deviation of reduced density from unity
\begin{equation}
\log \nu_{Ar}=3.39\log(\rho_r-1)+0.91,
\label{lohAr}
\end{equation}
where $\rho_r=\rho/\rho_c$,  $\rho_c$ is the critical density. The power index $3.39$ agrees with the corresponding theoretical value \cite{Beysens1987} $3.673$, which is known as slightly overestimating the real situation. The same behaviour is noted for subcritical fluctuations in liquid krypton and xenon calculated using the data \cite{NIST}.

It is important to note that the dependence in double-logarithmic co-ordinates for neon sufficiently deviates from linearity (see Fig.~\ref{nufig}). However, there is no any liquid-liquid structural transition located in this region, which could demonstrate such a behaviour \cite{Putintsev2001}. Thus, we need to explore the dependences of the speed of sound and heat capacities on the density, which are used in Eq.~(\ref{nu}), to reveal the source of the error.

\begin{figure}
\includegraphics[width=\textwidth]{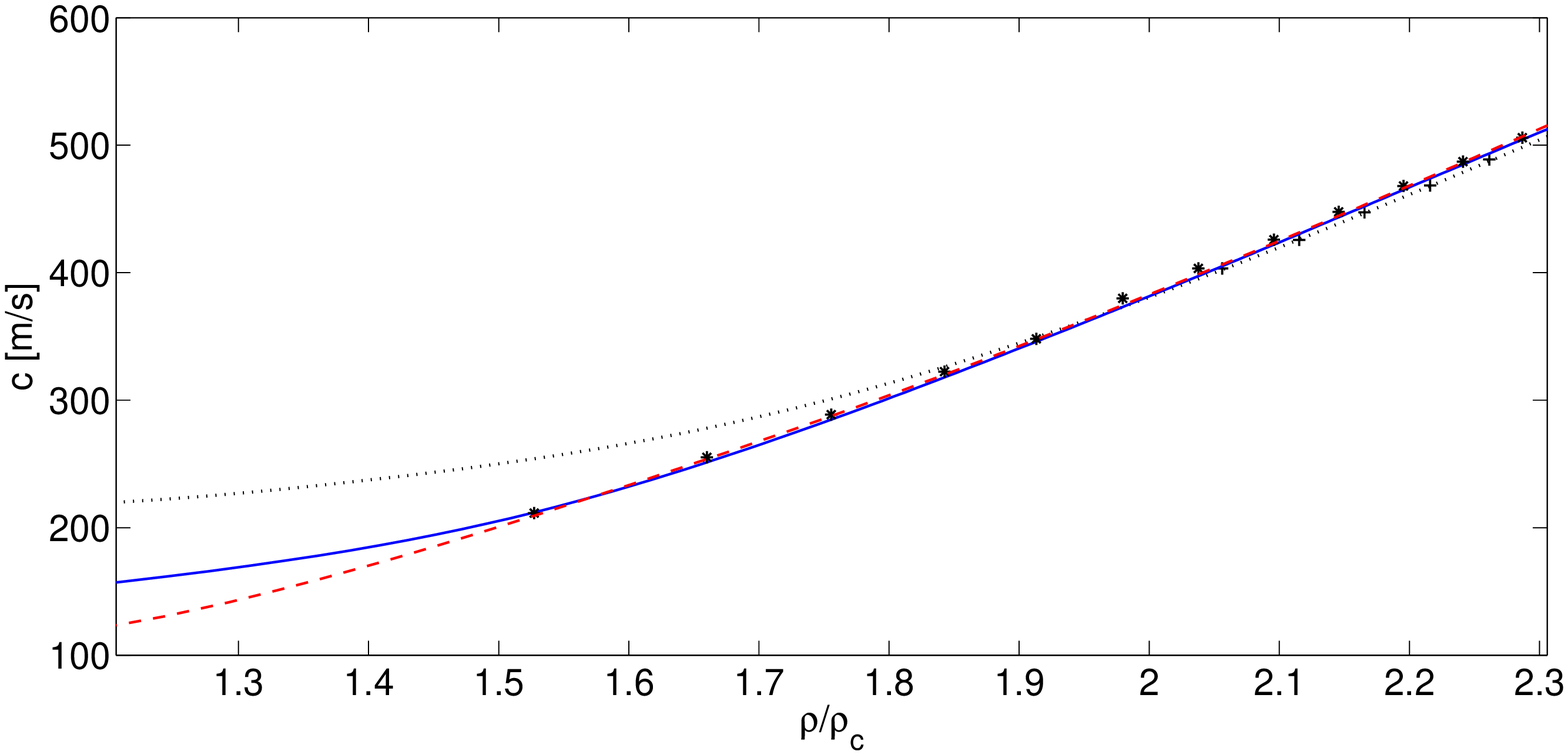}
\caption{Experimental and experimental-based values taken from \cite{Naugle1972} (pluses) and \cite{Gladun1966} (asterisks) as well as the fits $c=A_0+A_1T+A_2T.^2$ with $A_0=786.65~ms^{-1}$, $A_1=1.9754~ms^{-1}K^{-1}$, $A_2=0.33342~ms^{-1}K^{-2}$ (proposed in \cite{Naugle1972} , dotted line; it satisfies also the data from \cite{Rabinovich1988book}), \texttt{NIST Chemistry WebBook}\cite{NIST} (solid line) for neon. The dashed line corresponds to the scaled as $c=0.78c_{Ar}$ speed of sound in saturated liquid argon \cite{NIST} for the same interval of the dimensionless reduced density.}
\label{sos}
\end{figure}

Fig.~\ref{sos} shows various available experimental values of the speed of sound in liquid neon along the coexistence line and the known approximation proposed in \cite{Naugle1972}:
\begin{equation}
c=A_0+A_1T+A_2T.^2,
\label{capp} 
\end{equation}
where $A_0=786.65~ms^{-1}$, $A_1=1.9754~ms^{-1}K^{-1}$, $A_2=0.33342~ms^{-1}K^{-2}$. It satisfies also the data from \cite{Rabinovich1988book}.

One can see that all data are well agreed up to $37~K$ ($\rho_r=2.06$), the last point of direct measurements evaluated in the work\cite{Naugle1972}. In addition, the quadratic fit (\ref{capp}) proposed there coincides with the experimental-based data\cite{Gladun1966} and both continual approximations \cite{Rabinovich1988book,NIST} up to $39~K$ ($\rho_r=1.92$). 

However, the divergence between the fits \cite{Naugle1972,Rabinovich1988book} and the results\cite{NIST,Gladun1966} grows within the interval $43~K>T>39~K$ ($1.52<\rho_r<1.92$). This can be explained by the fact that the parabolic function of temperature is best adjusted to the reproduction of the speed of sound within the temperature interval $37~K>T>25~K$, where the actual measurements have been made. Its extrapolation to the practically subcritical region (as it is used in  \cite{Rabinovich1988book}) does not have a foundation.

On the other hand, NIST computational output is based on the experimentally verified values\cite{Gladun1966} for the region $43~K>T>39~K$ ($1.52<\rho_r<1.92$) and, therefore should be considered as confirmed within the mentioned interval. Table~\ref{errors} represents the relative errors for neon \cite{NIST} and argon (the rescaled data \cite{NIST}). 

The concluding region of discussion is the last $1.5~K$ before the critical point. As far as we know, there are no data of direct measurements within this interval. Therefore, the quality of approximation proposed in\cite{NIST} can be estimated only indirectly. Note that the speed of sound, the plot of which is shown as a solid line in  Fig.~\ref{sos} decays with decaying reduced density much more slower for $T>43~K$ than for $T<43~K$. The dotted line representing the fit based on the smaller temperatures exhibits the same qualitative behaviour. To check the supposition that the approximation \cite{NIST}, which provides a retarded decay of the speed of sound in neon, is the error source, we consider the behaviour of values of the speed of sound in argon\cite{NIST}, which are confirmed by a larger number of experimental data and are more warrant.

Such an approach is based on the facts of critical universality\cite{Beysens1987}, which demonstrate a coincidence of thermodynamic parameters for substances having the similar structure and the character of intermolecular interactions up to dimensionless scaling factors, and the theory of thermodynamic similarity\cite{Skripov2006book} based on these facts. As a confirmation, one can see that the rescaled speed of sound in saturated liquid argon (dashed line in Fig.~\ref{sos}) fits the known experimental data for neon with a high accuracy.In addition, the approximation (\ref{cappr}) for the speed of sound is even better within the interval $43~K>T>39~K$ ($1.52<\rho_r<1.92$) than NIST's fit (solid line) due to the smaller curvature in comparison with the latter, see Table~\ref{errors}. Whence, the faster decay of the speed of sound in neon given by the rescaling of argon's values for $T_c>T>39~K$ ($1<\rho_r<1.92$) could be argued as more physically realistic than the proposed by NIST.

\begin{figure}
\includegraphics[width=\textwidth]{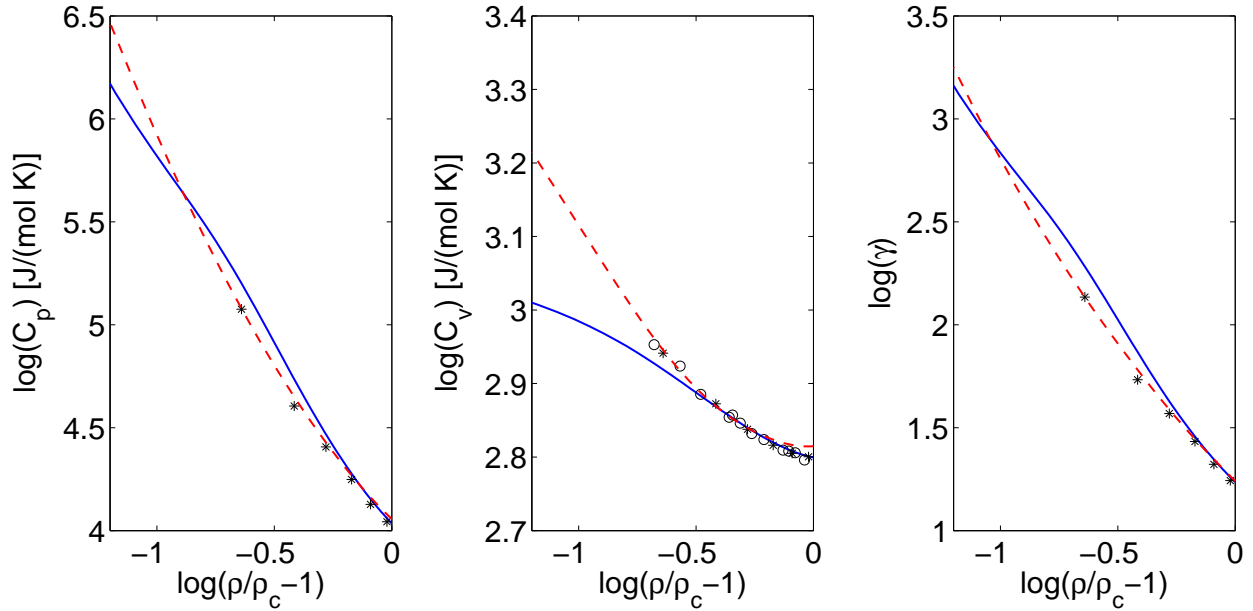}
\caption{The isobaric (left), isochoric (center) heat capacities and the heat capasity ratio (right) given by NIST (blue line) and measured in\cite{Gladun1966}: raw (circles) and processes (asterisks). The dashed line presents NIST data for argon multiplied by $0.93$.}
\label{figCpCv}
\end{figure}

Now let us consider the isobaric and isochoric heat capacities, the ratio of which is also used in (\ref{nu}). The corresponding plots are presented in Fig.~\ref{figCpCv}. One can see that all data provided by NIST in the temperature interval from $40~K$ up to the critical point demonstrate significant discrepancies. This conclusion follows at first from their deviation from the experimental data \cite{Gladun1966}, and at second from the shapes of curves represented in a logarithmic scale. As for the latter, the general theory of critical phenomena \cite{Beysens1987} claims the strict power-law divergence of the heat capacities approaching the critical point.
Therefore, the plots should be straight lines in logarithmic co-ordinates. But the solid lines do not satisfy this condition (Fig.~\ref{figCpCv}): the isobaric heat capacity curve weakly oscillates around the straight line, and the isobaric heat capacity not only deviates from the basic experimental data but even slows its growth approaching the critical region. Table~\ref{errors} shows the permanent growth of the relative errors for both heat capacities as the temperature approaches the critical one.  As a result, the plot of heat capacities ratio in Fig.~\ref{figCpCv} has a ``hump'', which corresponds to deviating line in the plot representing the reduced fluctuations (Fig.~\ref{nufig}).

On the other hand, it is possible to repeat the rescaling procedure for NIST’s data on argon with the goal to obtain a better approximation for both heat capacities of neon. The corresponding dependencies $C_{p,V}=0.93C_{p,V(Ar)}$ are shown in Fig.~\ref{figCpCv} as dashed lines. They quite well coincide with the real experimental data\cite{Gladun1966} within the interval $43~K>T>40~K$  see Table~\ref{errors} as well as keep the linear character in logarithmic co-ordinates, i.e. power-law divergence, when tend to the critical point.

\begin{table}
\label{errors}
\caption{Relative errors of approximation of the isobaric, isochoric heat capacities and the speed of sound with respect to the data from \cite{Gladun1966}. The reference \cite{NIST} marks the usage of direct data provided by \texttt{NIST Chemistry WebBook}for neon, and [Ar-corr.] corresponds to the correction proposed in the present work using rescaled NIST's data for argon.}
\begin{tabular}{ccccccc}
\hline
T,&$\varepsilon_{C_p}$ \cite{NIST},&$\varepsilon_{C_p}$ [Ar-corr.],&
$\varepsilon_{C_v}$ \cite{NIST},&$\varepsilon_{C_v}$ [Ar-corr.],&
$\varepsilon_{c}$ \cite{NIST},&$\varepsilon_{c}$ [Ar-corr.],
\\
K&\%&\%&
\%&\%&
\%&\%
\\
\hline
40&	0.8&	-1.7&0.0&	-0.7&-0.9&	0.5\\
41&	-0.6&	-1.9&0.4&	-0.1&-1.6&	0.3\\
42&	-8.5&	-3.8&0.9&	0.4&-0.5&	0.2\\
43&	-10.5&	0.1&2.5&	-0.3&-1.1&	0.8\\
\hline
\end{tabular}

\end{table}

\section{Conclusion}

The present study shows that the calculational approximation, which is realized in the database \texttt{NIST Chemistry WebBook} \cite{NIST} based on the algorithm \cite{Katti1986} can be considered as reliably for the acoustic and thermophysical data for liquid neon along the coexistence curve up to $40~K$ only. However the results \cite{NIST} sufficiently deviate from the actual experimental and physically expectable (in particular, from the point of view of a general theory of general critical phenomena) data within the subcritical region $40-44.49~K$. Therefore, the fit adjusted to the states far from the critical point can not be extrapolated to its vicinity. In addition, this region covers the temperature interval of about a few Kelvins. Whence, small deviations of the temperature used as an independent variable during its fitting may result in large deviations in large deviations of other state parameters.

On the other hand, the density varies sufficiently over this small interval of the temperature variations. An additional amplification of the sensitivity is provided by the application of the reduced volume fluctuation approach \cite{Goncharov2013}.

Thus, the processing of thermodynamic parameters from NIST database included into (\ref{nu}) shows that they all require corrections for $T>40~K$. But this correction can be easily evaluated via the corresponding rescaling (the coefficients are given in the work) of NIST’s data for liquid argon.

The work is supported by the grant No.~1391 of the Ministry of Education and Science of the Russian Federation within the basic part of research funding No. 2014/349 assigned to Kursk State University.


\end{document}